\def\real{{\tt I\kern-.2em{R}}}
\def\nat{{\tt I\kern-.2em{N}}}

\def\realp#1{{\tt I\kern-.2em{R}}^#1}
\def\natp#1{{\tt I\kern-.2em{N}}^#1}
\def\hyper#1{\ ^*\kern-.2em{#1}}

\def\st#1{{\tt st}(#1)}
\def\hyperreal{{^*{\real}}}
\def\hyperrealp#1{{\tt ^*{I\kern-.2em{R}}}^#1} 

\def\hypernatp#1{{{^*{{\tt I\kern-.2em{N}}}}}^#1} 
\def\eskip{\hskip.25em\relax}

\def\Hyper#1{\hyper {\eskip #1}}
\def\leaderfill{\leaders\hbox to 1em{\hss.\hss}\hfill}
\def\srealp#1{{\rm I\kern-.2em{R}}^#1}

\def\pars{\par\smallskip}
\def\parm{\par\medskip}

\def\b#1{{\bf #1}}
\def\ref#1{$^{#1}$}

\def\m@th{\mathsurround=0pt}
\def\rightarrowfill{$\m@th \mathord- \mkern-6mu \cleaders\hbox{$\mkern-2mu 
\mathord- \mkern-2mu$}\hfil \mkern-6mu \mathord\rightarrow$}
\def\leftarrowfill{$\mathord\leftarrow
\mkern -6mu \m@th \mathord- \mkern-6mu \cleaders\hbox{$\mkern-2mu 
\mathord- \mkern-2mu$}\hfil $}
\def\noarrowfill{$\m@th \mathord- \mkern-6mu \cleaders\hbox{$\mkern-2mu 
\mathord- \mkern-2mu$}\hfil$}
\def\orgate{$\bigcirc \kern-.80em \lor$}
\def\andgate{$\bigcirc \kern-.80em \land$}
\def\inverter{$\bigcirc \kern-.80em \neg$}

\magnification=\magstep1
\tolerance 10000
\baselineskip  14pt
\hoffset=.25in
\hsize 6 true in
\vsize 8.5 true in
\centerline{\bf An Operator Equation and Relativistic}
\centerline{{\bf Alterations in the Time for Radioactive Decay}}   
\bigskip
\centerline{{\bf Robert A. Herrmann}\footnote*{Partially funded by a grant 
from the United States Naval Academy Research Council. This version corrects notational errors that appear in the published version [8], adds new material as well as presenting a simple and correct derivation for the alteration in the rate of radioactive decay. }}
\centerline{Mathematics department}
\centerline{United States Naval Academy}
\centerline{572C Holloway Rd.}
\centerline{Annapolis, MD 21402-5002}
\vskip 24pt
\noindent ABSTRACT. In this paper, using concepts from the nonstandard 
physical world, the linear effect line element is derived. Previously, this 
line element was employed to obtain, with the exception of radioactive decay, 
various experimentally verified special theory relativistic alterations in 
physical measures.  This line element is now used to derive, by means of 
separation of variables, an expression that predicts the same increase in the  
decay time  for radioactive  material as that predicted by the Einstein 
time dilation assumption. This indicates that such an increase in lifetime can 
be attributed to an interaction of the radioactive material with a nonstandard photon-particle medium so as to maintain hyperbolic velocity-space behavior.\pars 
\par\bigskip\smallskip \noindent {\bf 
1.\hskip 1.25em Introduction.} \parm

 In [7], a specific operator equation 
related to a partial differential equation, the Schwarzschild and linear 
effect line elements, and the concept of separation of variables are used to 
derive various relativistic alteration expressions. As discussed in [6], the 
measurable aspects of the special theory of relativity need not be related to 
the so-called general time dilation or length contraction concepts. It is conjectured that all 
of the special theory predicted alterations in measured quantities can be 
obtained by means of electromagnetic propagation model. The quantities 
must be measured by infinitesimal light-clocks or an equivalent approximating device and, in order to avoid the 
model theoretical error of generalization, no other instrumentation is 
allowed. \pars

 In [7], with the exception of relativistic alterations in 
the lifetime of radioactive material, all of the major special theory 
relativistic alterations in measured quantities are obtained from the linear 
effect line element. What is derived first is a statement relating 
infinitesimal time measurements, where these measurements are interpreted as 
changes in the number of counts or ``ticks'' that occur within a measuring 
infinitesimal light-clock. The use of infinitesimal light-clocks to obtain the experimentally verified alterations 
indirectly implies that such alternations may be caused by (emis) - interactions within a nonstandard photon-particle medium (NSPPM) using a photon particle behavioral model. \pars 

The main purpose of this paper is to present a derivation of the 
linear effect line element using a rather simple infinitesimal NSPPM process and to use the methods in 
[7] to predict the experimentally verified [1] relativistic 
alteration in radioactive decay time. \parm 

\noindent {\bf 2.\hskip 1.25em The Linear Effect Line Element.} \parm

 In what follows, we are considering a 
standard set-theoretic superstructure ${\cal M} = \langle \real,\in, = \rangle$ 
with ground set the real numbers $\real$ and this structure is embedded into a 
nonstandard elementary extension $\Hyper {\cal M}= \langle \hyperreal, \in, = 
\rangle$ that is an enlargement [10].  In [5a, b], the chronotopic 
interval (i.e. proper time interval) is derived and yields the following 
interpretation. An infinitesimal light-clock is in linear and uniform motion, 
and $L$ is an infinitesimal number representing the length of the arm. 
The infinitesimal light-clock has  ``ticked'' $\Pi_m\in \nat_\infty$ times, 
where $\nat_\infty$ is the set of Robinson infinite natural numbers. The 
apparent physical world linear distance traversed by the electromagnetic 
radiation in its to-and-fro motion within the moving infinitesimal light-clock 
is $\st{2L\Pi_m}.$\pars 

At a space-time point, there are four infinitesimal 
light-clocks. These four infinitesimal light-clocks are used to obtain Einstein 
measurements [9], which are obtained by the ``radar'' method for determining the relative velocity of a moving light-clock, the distance and ``time'' measurements at a point $m$ in motion relative to a point $s.$ The radar method is used since retaining this mode of measurement the twin anomaly is eliminated. This anomaly is not eliminated when other derivations, independent from the mode of measurement, are used. Further, the GR approach does not eliminate this anomaly under reasonable conditions and, especially, when three objects are used [9]. In general, Einstein time, distance and velocity are $t_E = (1/2)(t_3 + t_1),\ r_E = (1/2)c(t_3 - t_1),\  v_E = r_E/t_E,$ when defined. The $t_1$ is the moment of time at the $s$-point when the light-pulse is emitted and $t_3$ the time when it is received back at the $s$-point after being ``reflected'' from the $m$-point. Notice that when $r_E = 0$, then $t_E = t^3.$ \pars

Three infinitesimal light-clocks that correspond to a Cartesian coordinate system are used to measure local distance at the $s$-point and the fourth infinitesimal 
light-clock is used to measure time at the $s$-point. In particular, the superscript and subscript $s$ represents local measurements about the $s$-point, using various devices, for laboratory standards (i.e. standard behavior), and infinitesimal light-clocks or approximating devices such as the atomic-clocks. [Due to their construction, atomic clocks are affected by relativistic motion and gravitational fields approximately as the infinitesimal light-clock's counts are affected.] Superscript or subscript $m$ indicates local measurements at the $m$-point for an entity considered at the $m$-point in motion relative to the $s$-point. The local time measure at the $m$-point is the Einstein time, via the radar method, as registered at the $s$-point. The relative velocity and distance between the points are also the Einstein measurements. These measurements are used to investigate $m$-point behavior. To determine how physical behavior is being altered, the $m$-point and $s$-point measurements are compared. \pars

Let $\omega \in \nat_\infty.$ Since the sequence $S_n = n \rightarrow \infty,$ then as 
proved in [4, p. 100] for each $r \in \real,$ there exists an $x \in \{m/\omega
\mid (m \in \Hyper {\b Z})\land (\vert m\vert < \omega^2)\}$ such that $x 
\approx r.$  
 As discussed in [6] and due to this result,  
the use of infinitesimal light-clocks allows for time and 
length measurements to be considered as varying over intervals of real 
numbers. This might be expressed by saying that they can take on a ``continuum'' 
of values. Thus standard analysis can be applied.\par
Let $(x^s,y^s,z^s,t^s)$ be Cartesian coordinates that are the 
standard part of $s$-point infinitesimal light-clock 
measurements for locally stationary points. Let $(x^m,y^m,z^m,t^m)$ be Cartesian coordinates that are the 
standard part of $m$-point infinitesimal light-clock 
measurements as registered at the $s$-point via Einstein measures since this application is for special relativity. At the $s$-point the infinitesimal light-clock count (``ticks'') are used to obtain time intervals. At the moment when the $s$-point infinitesimal light-clock has count number $\Pi_s'$, a signal is sent to the $m$-point and its reception, as measured by Einstein time at the $s$-point, is at count number $\Pi'_m$. A second single is sent from the $s$-point to the $m$-point and the two count numbers are $\Pi_s'',\ \Pi_m''.$ The infinitesimal light-clock measured interval is $\Pi_s = \Pi_s'' - \Pi_s'.$ The length of 
$s$-point time interval is the real number $\st {u\Pi_s}= \Delta t^s.$ For the $m$-point, even when using Einstein measures, the length of the time interval can be expressed as $\Pi_m =\Pi''_m - \Pi'_m$, and has the real value $\st {u\Pi_m}.$ Using these count numbers, the expression derived in [5a,b] is
$$(\st {L\Pi_m})^2 = (\Delta t^s)^2 c^2 -((\Delta x^s)^2 +(\Delta y^s)^2 +(\Delta z^s)^2),\eqno (2.1)$$
where $c$ is the to-and-fro local measurement at the $s$-point for the velocity of electromagnetic 
radiation. Physical measurements can be considered as absolute with respect to a third position fixed in the NSPPM. In this case, the three relative velocities are related by the velocity composition expression [5b, p. 48], where the NSPPM velocities follow simple linear Galilean rules.\pars

From a quantum-physical viewpoint, certain photon interactions can be viewed as mimicking  the basic to-and-fro light-clock process. Hence, light-clocks are the best form of timekeeping that satisfies the rules for infinitesimalizing. This allows for the infinitesimalizing of expression
(2.1) and yields $$dS^2 = (dt^s)^2 c^2 -((dx^s)^2 +(dy^s)^2 +(dz^s)^2).\eqno (2.2)$$     
Equation (2.2) are infinitesimal light-clock measures. This correspond to the classical approach, when $ds$ is restricted to infinitesimal light-clock counts. Expression (2.2)
is that which is used when (2.1) is extended so as to include nonuniform 
behavior. 
We often write $(dr^s)^2 = (dx^s)^2 +(dy^s)^2 +(dz^s)^2$. \pars

Expression (2.2) relates behavior of infinitesimal light-clock counts but
does not indicate that there might be a cause for such behavior. Distinct from 
the classical approach to the special theory, a cause can be postulated 
relative to the behavior of the NSPPM. 
In what follows, the timing infinitesimal light-clocks are used as 
an analogue model to  
investigate how the NSPPM behavior is altered  
by a 
process $P$. It is shown [5, pp. 50-54] that, for special relativity, global NSPPM light propagation within our physical world using radar measured velocities is related to an NSPPM  hyperbolic velocity-space. The $P$-process within the NSPPM has the effect that the Einstein measurements for relative velocity $v$ are directly related to an NSPPM velocity $w = (c/2)\ln {((1+v^2/c^2)/(1 - v^2/c^2))}$ [5b, p. 51. Further, the $P$-process is related to motion 
viewed from  points in  Euclidean  
space. Suppose that the counts of these measuring infinitesimal light-clocks 
are affected by 
this $P$-process. We seek a relationship $\phi(dx^m,dy^m,dz^m,dt^m)$ or 
$\Phi(dr^m,dt^m)$ between altered counts 
as it is obtained by measuring infinitesimal light-clocks 
so that $L\Pi_m = \phi(dx^m,dy^m,dz^m,dt^m)$ or  $L\Pi_m = \Phi(dr^m,dt^m).$  
 \pars

Following [5, 6], let $v,d$ and $c$ behave within a monadic neighborhood as if 
they 
are constant with respect to $P.$ Since behavior in an infinitesimal neighborhood is supposed to be simple behavior, then the  simple Galilean velocity-distance law holds. 
For photon behavior with a moving source velocity $v+d$, $$((v+d) + c)dt^s = (v+d)dt^s + cdt^s = dR^s + dT^s,$$ $$dT^{s} = cdt^{s},\ dR^s =(v +d)dt^s,\ dt^s\not= 0, \ {{dR^{s}}\over{dT^s}} = {{v+d}\over{c}},\ .\eqno (2.3)$$ 
Suppose that, due to $P$, 
a smooth microeffect [3] alters the infinitesimal light-clock counts. 
For $(dr^m)^2 = (dx^m)^2 +(dy^m)^2 +(dz^m)^2$, this alteration is 
characterized by the infinitesimal linear transformation
(A): $dr^{s} = (1-\alpha\beta) dr^{m} -  \alpha dT^{m},$   
(B): $dT^{s} = \beta dr^{m} + dT^{m},\  dT^{m} = cdt^{m},$ 
where $\alpha, \ \beta$ are to be determined. Since the effects 
are to be observed in the physical world, $\alpha, \ \beta$ have 
standard values.
 \pars

Substituting (A) and (B) into (2.2) yields 
$$dS^2 = (1-\alpha^2)(dT^{m})^2 + 2(\alpha +\beta(1-
\alpha^2))dr^{m}dT^{m}+$$
$$(\beta^2 -(1-\alpha\beta)^2)(dr^{m})^2. \eqno (2.4)$$\par
The simplest real world aspect of time interval measurement that assumes that 
timing counts can be added or subtracted is 
transferred to a monadic neighborhood and requires  
$dT^{m}$ to take on positive or negative infinitesimal values. 
As done for space-time, suppose that the 
$P$-process is symmetric with respect to the past and future sense of a time variable. This implies that $dS^2$ is unaltered when $dt^m$ is replaced by 
$-dt^m$. This implies that the transformation needs to be restricted so that $2(\alpha + \beta(1-
\alpha^2)) = 0.$  For simplicity of calculation,  let $\alpha = -\sqrt{1 - \eta}.$ Hence, 
$\beta = \sqrt {1 - \eta}/\eta.$ Substituting into (A) and (B) yields
$$dr^{s} = {{1}\over{\eta}}dr^{m} + \sqrt{1-\eta}\, dT^{m}$$
$$dT^{s}= {{\sqrt{1-\eta}}\over{\eta}}dr^{m} + dT^{m}. \eqno (2.5)$$
Combining both equations in (2.5) produces 
$${{dr^{s}}\over{dT^{s}}}=
{{{{1}\over{\eta}}{{dr^{m}}\over{dT^{m}}} +\sqrt{1-\eta}}\over
{{{\sqrt{1-\eta}}\over{\eta}}{{dr^{m}\over{dT^{m}}}+ 1}}}. \eqno (2.6)$$\pars 
The $x^m,\ y^m, z^m$ are not dependent upon $t^m$. Hence the (total) derivative 
$dr^{m}/dT^m= 0$. Using (2.6), $dr^{s}/dT^{s} = (v+d)/c 
= \sqrt {1- \eta}$ or $\eta = 1-(v+d)^2/c^2=\lambda\not=0.$ 
We note that using $\alpha = \sqrt {1 - \eta}$ yields the 
contradiction $(v+d)/c < 0$ for the case being considered that $0 \leq v+d < c.$ 
By substituting $\eta$ into (2.5) and then (2.5) into (2.2) (see 
(2.4)), we 
have, where $dT^m = cdt^m,$ the linear effect line element
$$(cdt^s)^2 - (dr^s)^2 = dS^2 = \lambda(cdt^{m})^2 - (1/\lambda)(dr^{m})^2. \eqno (2.7)$$
This linear effect line 
element yields a special theory line element if $v = v_E$ (the Einstein 
measure of the relative velocity)  
and correlates 
special theory effects to a describable 
$P$-process. This process includes the appropriate linear Galilean rules for uniform motion that in the NSPPM yield the hyperbolic velocity-space for the observed physical-world. Equation (2.7) represents the necessary alteration 
in the infinitesimal light-clock counts when they are 
affected by the $P$-process. Since the line 
element is a relation between infinitesimal quantities, the expression  
``linear effect'' is not intended to imply that the (standard) path of motion 
of the analogue infinitesimal light-clock is necessarily linear. \par

[Transforming (2.2) via the spherical coordinate transformations, a similar argument for a static $P$-process yields a general radial effect line element. By selecting various ``potential'' velocities $v$ and $d$ (i.e. whatever $v$ and $d$ are they have velocity units of measure), this general radial line element is transformed, at the least, into the Schwarzschild, Schwarzschild with cosmological constant, de Sitter, and even the Robertson-Walker line elements. Indeed, with appropriate $v$ and $d = 0$, (2.7) is the approximating Newtonian line element. Moreover, (2.7) applies directly to regions where gravitational potentials are constant or approximately so. For gravitational potentials, Einstein measurements are not used.]\pars 
 
Following the usual practice for decay purposes, it is assumed that the atomic 
structure is momentary at rest when decay occurs. This is modeled by letting
$dr^s = 0$ in (2.2) and $dr^m = 0$ in (2.7). Hence we have that (*) $\gamma dt^m = 
dt^s,$ where $\gamma = \sqrt \lambda\not= 0.$  This does not indicate a change in
the concept of time. As implied in [6] and shown in [5], for zeroed infinitesimal light-clocks, $dt^m = 
u\Delta\Pi_m,\ dt^s = u\Delta\Pi_s.$  The NSP-world time unit $u$ is not altered, but 
the infinite count number $\Delta \Pi_s$ has been altered. 
Using this analogue 
approach, if the verified alterations in decay 
rates  can be predicted, then this would indicate that radioactive decay itself is related to an (emis) effect.  \parm\vfil\eject

\noindent {\bf 3.\hskip 1.25em The Derivation of the Decay Prediction.} \parm

The method of separation of variables as used in [7] is the consistent
 underlying procedure used to predict the verified alteration
in decay time. Unfortunately, in [7] a confusing typographical error 
occurs in expressions (3.2) and (3.3). The symbols $h(r^s,t^s)$ should be 
replaced with $h(r^s)$ and $H(r^m,t^m)$ should be replaced with $H(r^m).$  
\pars

Let $N(t^s)$ denote a measure for the number of active entities at the 
light-clock count time $t^s$ and $\tau_s$ be the 
(mean) lifetime. These measures are taken within a laboratory and are used as 
the standard measures. This is equivalent to saying that they are, from the 
laboratory viewpoint, not affected by relativistic alterations.    
The basic statement is that there exists some $\tau \in (0,B]$ such that (*) $(-\tau)dN/dt = N.$ Even though the number of active entities is a natural number, this 
expression can only have meaning if $N$ is differentiable 
on some time interval. But, since the $\tau$ are averages and the number of entities is usually vary large, then such a differentiable function is a satisfactory approximation.  
Recall that the required operator expression is $$D(T) = k (\partial/\partial t)(T).\eqno (3.1)$$ \pars 

Let $k = 1$ and $h(r) = 0\cdot r^2 + 1 = 1.$ Then define $T(r,t) = h(r)N(t) = (0\cdot r^2 + 1){N}(t),$ where $r^2 = x^2 + y^2 + z^2,$ and let $D$ be the identity map $I$ on $T(r,t).$ Then $D(T(r,t))= D(h(r))N(t) = D(0\cdot r^2 +1)N(t) = 1\cdot N(t)$ and, in this form, $D$ is considered as only applying to $h$ and it has no effect on $N(t).$ In this required form [5b, p. 62; 7, p. 60], first let $r= r^s$ and $t = t^s.$  Further, consider $T(r^m, t^m) = H(r^m)\overline{{N}}(t^m), \ H(r^m) = 0\cdot (r^m)^2 + 1 = 1.$ In order to determine whether there is a change in the $\tau_s$, one considers the value $N(t^s) = \overline{N}(t^m).$ This yields the final requirement for $T$. Notice that $t^m$ is Einstein time as measured from the $s$-point. This is necessary in that the $v = v_E$, which is a necessary requirement in order to maintain the hyperbolic-velocity space behavior of $v$. \pars 
Applying (3.1) to $T$ and considering the corresponding differential equation (*) and the chain rule, one obtains  that there exist a real number $\tau_s$ such that
  
$${N}(t^s) = (-\tau_s)(d/dt^s){N}(t^s)=$$ $$(-\tau_s) (d/dt^m)\overline{{N}}(t^m) (d/dt^s)(t^m)=$$ $$ (-\tau_s/\gamma)(d/dt^m)\overline{{N}}(t^m).\eqno (3.2)$$
\noindent And, with respect to $m$, and for $\tau_m$ 
$$\overline{{N}}(t^m)= (-\tau_m)(d/dt^m)\overline{{N}}(t^m). \eqno 3.3)$$\par
Using (3.2) and (3.3) one obtains that ${\tau_m} = \tau_s/\gamma.$ (In the linear effect line element $d = 0.$) This is one of the well-known expressions for the prediction for the alteration of the decay rates due to relative velocity (that is $v_E$).  The $\tau_s$ can always be taken as measured at rest in the laboratory since the relative velocity of the active entities is determined by experimental equipment that is at rest in the laboratory. 
\parm

\centerline{\bf References}\par\medskip
\noindent 1. Bailey, J. and et al. Measurements of relativistic time dilatation for 
positive and negative
 muons in a circular orbit, {\it Nature}, 268(July 
1977), 301-305.\parm
\noindent 2. Herrmann, R.A. Supernear functions, {\it Math. 
Japonica}, 30(1985):125-129. \parm   
\noindent 3. Herrmann, R.A. Fractals and ultrasmooth microeffects, 
{\it J. Math. Phys.}, 30(April 1989):805-808.\parm
\noindent 4. Herrmann, R.A. {\it The Theory of Ultralogics}\hfil\break http://arXiv.org/abs/math.GM/9903081 and\hfil\break http://arXiv.org/abs/math.GM/9903082 \parm
\noindent 5a. Herrmann, R.A. {\it Constructing Logically 
Consistent Special and 
General
Theories of
Relativity}, Math. Dept., U.S. 
Naval Academy, Annapolis, MD, 1993. \parm
\noindent 5b. Herrmann, R. A. {\it Nonstandard Analysis Applied to the Special and General Theories of Relativity - The Theory of Infinitesimal Light-Clocks,} (1994 - )\hfil\break
http://arxiv.org/abs/math/0312189   \parm
\noindent 6. Herrmann, R.A. Special relativity and a nonstandard substratum, 
{\it Speculat. Sci. Technol.}, 17(1994):2-10. http://arxiv.org/abs/physics/0005031 \parm
\noindent 7. Herrmann, R. A., Operator equations, separation of variables and 
relativistic alterations, {\it Intern. J. Math. Math. Sci.}, 18(1)(1995):59-62.\parm
\noindent 8. Herrmann, R.A., An operator equation and relativistic alterations in the time for radioactive decay, {\it Intern. J. Math. Math. Sci.}, 19(2)(1996):397-402.\parm

\noindent 9. Proknovnik, S.J. {{\it The Logic of 
Special Relativity}, Cambridge University Press, 
London,
1967. \parm\end
\noindent 10. Stroyan, K.D. and W.A.J. Luxemburg, {\it Introduction to the Theory} of Infinitesimals},
Academic Press, New York, 1976.\parm

\end